\documentclass[graybox]{svmult}

\usepackage{type1cm}        
\usepackage{makeidx}         
\usepackage{graphicx}        
\usepackage{multicol}        
\usepackage[bottom]{footmisc}
\usepackage{comment}

\usepackage{newtxtext}       %
\usepackage[varvw]{newtxmath}       
\usepackage{url}
\usepackage{tabularx}
\newcolumntype{L}[1]{>{\raggedright\arraybackslash}p{#1}} 
\newcolumntype{C}[1]{>{\centering\arraybackslash}p{#1}} 
\newcolumntype{R}[1]{>{\raggedleft\arraybackslash}p{#1}} 

\usepackage{array}
\usepackage{amstext}
\usepackage{amsmath}

\makeindex             

\begin{document}

\title{Spatial Conceptual Modeling:\\Anchoring Knowledge in the Real World}
\titlerunning{Spatial Conceptual Modeling}

\author{Hans-Georg Fill\orcidID{0000-0001-5076-5341} }

\institute{Hans-Georg Fill \at University of Fribourg, Research Group Digitalization and Information Systems, 1700 Fribourg, Switzerland, \email{hans-georg.fill@unifr.ch}\\
This is a preprint of the following chapter: Fill, H.-G., Spatial Conceptual Modeling - Anchoring Knowledge in the Real World, published in Metamodeling: Applications and Trajectories to the Future - Essays in Honor of Dimitris Karagiannis, edited by Hans-Georg Fill and Harald Kühn, 2024, Springer. The final authenticated version is available online at: http://dx.doi.org/10.1007/978-3-031-56862-6}

%
%
\maketitle

\abstract{This paper introduces the concept of spatial conceptual modeling, which allows anchoring mental world knowledge in the physical world using augmented reality technologies. For a first formal characterization, we describe a mapping from the spatial information concepts location, field, object, network, and event, as used in spatial computing, to conceptual modeling concepts using the FDMM formalism. This allows to identify necessary adaptations at the metamodeling level to make the approach applicable to arbitrary types of spatial conceptual modeling languages. Finally, possible application areas of spatial conceptual modeling in the medical domain, manufacturing and engineering, physical IT architectures and smart homes, supply chain management and logistics, civil engineering, and smart cities and cultural heritage are discussed.}


\section{Introduction}
\label{sec:intro}

Conceptual modeling is a widely-used technique for representing aspects about the physical and social world in order to support human understanding and communication~\cite{mylopoulos1992,HarerF20}. Further, conceptual models can act as interfaces to digital technologies, thereby using the contained knowledge for the configuration of machines and leading to the realization of innovative IT applications~\cite{Fill20}. In recent years, there has been a growing interest in virtual reality, augmented reality, and mixed reality technologies~\cite{WohlgenanntSS20,pwc2022}, which are part of the broader field of \emph{spatial computing}. These technologies allow users to be fully immersed in virtual 3D environments through the use of special headsets (\emph{virtual reality}), or to project virtual information onto objects in the real world using see-through displays or smartphones (\emph{augmented reality}), or combinations thereof (\emph{mixed reality}).

The basic components of such applications have been available for a considerable period of time, and many of the technical concepts required have been well researched and developed to a high degree of maturity~\cite{schmalstieg2016}. However, technological advances in the hardware of headsets and mobile devices have dramatically simplified the usability of such devices. In addition, the cost of these devices has come down significantly. Today, wireless headsets such as the Microsoft HoloLens\footnote{https://www.microsoft.com/en-us/hololens}, the Meta Quest headsets\footnote{https://www.meta.com/quest/} or, in the near future, the Apple Vision Pro\footnote{https://www.apple.com/apple-vision-pro/} can be used out of the box, without having to connect them to powerful graphics workstations anymore. At the same time, the quality of headsets has improved over the years, affecting usability by reducing the potential for motion sickness and improving user acceptance~\cite{caserman2021}. 

As a result, it has become increasingly feasible to explore the use of these technologies in many business areas. These range today from applications in engineering, e.g.\ to support the assembly of machines~\cite{regenbrecht2005}, medical applications~\cite{munzer2019}, robot interactions~\cite{DelmericoPBOVCN22}, IT management~\cite{Crevoiserat23} to the gaming industry. While the technologies to compute and display the visual representations are an essential part of such applications, additional components are required to deliver the expected experience. 

The field of spatial computing studies these aspects on a more general level~\cite{ShekharFA16}. According to a broad definition given by Greenwold, spatial computing can be characterized as "\emph{human interaction with a machine in which the machine retains and manipulates referents to real objects and spaces}"~\cite{greenwold2003}. A central concept here is that such systems are aware of their location, be it in absolute terms such as position on the Earth, or in relative terms in the form of the distance to a reference point or origin. At a more granular level, location also includes orientation in space and how this is used to interact with a user. For example, in augmented reality applications, the user's position and orientation in space are used to compute overlays of the real world in the form of graphical information to augment the user's perception.

In the following, a synthesis of conceptual modeling with spatial computing will be described. This will be denoted as \emph{spatial conceptual modeling}. In contrast to previous techniques and tools in conceptual modeling that relied on paper-based, two or three-dimensional electronic formats, spatial conceptual modeling will permit to anchor the contents of conceptual models in the real world. This allows to attach the knowledge in these models to physical objects and/or position it spatially. For accomplishing this, augmented reality technologies can be employed to visualize these anchorings if necessary. It will be discussed, which changes this requires on the level of metamodeling, i.e.\ the conceptualization and technical implementation of modeling languages and methods. Further, use cases for spatial conceptual modeling will be illustrated.

The remainder of this chapter is structured as follows. In Section~\ref{sec:foundations}, foundations on conceptual modeling and metamodeling, as well as spatial computing and augmented reality will be presented. Subsequently, in Section~\ref{sec:spatial_conceptual_modeling}, the concept of spatial conceptual modeling will be elaborated and formally characterized using the FDMM formalism. In addition, potential use cases for spatial conceptual modeling will be illustrated. In Section~\ref{sec:conclusion} a conclusion and outlook on further research will be given.

\section{Foundations}
\label{sec:foundations}

For achieving a common understanding of the components of spatial conceptual modeling, a brief overview on the foundations of conceptual modeling and metamodeling, as well as spatial computing and augmented reality will be outlined in the following.

\subsection{Conceptual Modeling}

Conceptual modeling is concerned with the explicit representation of some aspects of the physical or social world around us~\cite{mylopoulos1992}. It is based on pre-defined elements or scripts that constrain what can be expressed in the models. In the sub-field of enterprise modeling for example, procedural knowledge about business processes, organizational knowledge, as well as knowledge about the enterprise architecture can be represented in this way~\cite{KaragiannisBB16,HinkelmannGKTMW16,SandkuhlFHKMOSU18}. This knowledge is typically codified using formal or semi-formal enterprise modeling languages~\cite{BorkF14}. In the past, a large range of frameworks and modeling methods have been developed for enterprise modeling, including for example the Business Process Management Systems (BPMS) Paradigm~\cite{karagiannis1996}, Multi-Perspective Enterprise Modeling (MEMO)~\cite{Frank14}, the Semantic Object Model (SOM)~\cite{FerstlS90}, or the 4EM Method~\cite{SandkuhlSPW14}.

Besides these academic approaches, some of which have been successfully deployed in industry, also a range of industrial approaches were proposed in conceptual modeling. This includes for example a large number of international standards, which lead to benefits for companies in terms of compatibility and repeatability~\cite{MoserBUK17}. Examples for such standards include ArchiMate~\cite{opengroup2023archimate}, BPMN~\cite{omg2024bpmn}, UML~\cite{omg2017uml}, or DMN~\cite{omg2023dmn}, which can be either used individually or in combination, e.g.~\cite{CurtyF22}. 

\subsection{Metamodeling}

Whereas conceptual models may also be created using pen and paper, any serious practical application requires today the use of IT-based modeling tools due to the complexity and size of the models. These tools not only permit to graphically represent the models but also to process them using algorithms and exchange them with third parties. Although such modeling tools may be created for one particular modeling language only, where the modeling primitives are hard-coded, the continuous evolution of modeling standards and languages would lead to much effort in their adaptation. In addition, conceptual modeling languages can be customized for specific purposes, or entirely new domain-specific languages can be developed to ensure optimal coverage of domain and user requirements~\cite{karagiannis2016domain,karagiannis2022domain}.

For these reasons, so-called metamodeling-based approaches have been designed. These correspond to typical approaches in knowledge representation and knowledge-based systems~\cite{studer1998knowledge,karagiannis1994,mylopoulos1992}, where a \emph{metamodel} acts as a terminological component (TBox), resulting from an iterative knowledge acquisition effort~\cite{karagiannis2015agile}. The metamodel thus formally defines the modeling language in such a way that it can be easily adapted if needed. This is in contrast to traditional approaches in compiler construction and language processing where the lexical and syntactic analysis as well as the actual code generation have to be explicitly specified~\cite{wirth1996compiler}. In metamodel-based approaches, common abstractions of typical metamodels are provided which are used to define an individual metamodel. These are denoted as the \emph{meta-metamodel}. These abstractions act as axioms and include for example concepts such as classes, relationclasses, attributes, or diagram types. When creating a metamodel, it is being reverted to these axioms for defining the terminological component. Metamodeling platforms - such as ADOxx for example~\cite{FillK13} - can then interpret the metamodel based on the axioms, whose semantics are hard-coded in the platform~\cite{KaragiannisK02}. In this way, the platforms can generate model editors for the specified modeling language. The created model instances then act as the assertion component (ABox) which contain the actual knowledge to be represented.

The main advantage of using metamodeling-based approaches and metamodeling platforms is the increased productivity in developing metamodels and thus new conceptual modeling languages. This is due to the pre-implemented concept interpretation functionality via the meta-metamodel in the terminology component. It eliminates the need for re-implementation and simplifies the creation of corresponding model editors and model processing environments.~\cite{karagiannis1994,karagiannis2015agile}. 

\subsection{Spatial Computing}
\label{subsec:spatial_computing}

Although a lot of research in \emph{spatial computing} has been done mainly in the area of geographic information systems, it can also be viewed from a broader perspective~\cite{ShekharFA16}. According to Kuhn and Ballatore, spatial computing can be characterized by \emph{spatial information} and \emph{spatial computations}~\cite{KuhnB15}. In the following these will be briefly summarized. For further details it is being referred to the original source~\cite{KuhnB15}. 

Kuhn and Ballatore further classify the properties of spatial information into the five core content concepts \emph{Location}, \emph{Field}, \emph{Object}, \emph{Network}, and \emph{Event} and two core quality concepts \emph{Granularity} and \emph{Accuracy}. Thereby, location as the most fundamental concept is regarded as a \emph{relation}. The location is therefore always determined relative to something else, e.g.\ in a coordinate system relative to an origin. The field concept, which originates from physics permits to describe phenomena in a space of interest by a single value of an attribute. This is done by mathematical functions that map positions in space to values. An example would be a temperature field where each position in space is assigned a temperature value through a mathematical function. The object concept captures individual things that extend in space, including physical, mental, or social entities. Objects have an identity for tracking their properties and relations over time. The network concept is used to establish connections between objects. Networks thereby correspond to mathematical graphs, i.e.\ including nodes and edges, which may thus be used for computations, e.g.\ to determine the shortest path between two objects. The event concept refers to the temporal aspect in spatial information. It can have relations to fields, objects, and networks. The quality concept of granularity relates to the level of detail or precision, which is used for expressing spatial information, while accuracy refers to whether something is described correctly given a particular granularity.

In addition to the concepts for spatial information, Kuhn and Ballatore provide primitives for spatial computing operations, which can be combined for more complex computations. These are attached to spatial information content concepts and include for example topological operations such as \emph{isAt} and \emph{isIn} for locations to determine whether something is in contact \emph{with} or contained \emph{in} some other entity, as well as temporal operations such as \emph{when} to determine the date of an event.

\subsection{Augmented Reality}

Augmented reality (AR) is a technique for superimposing virtual information such as visual media, audio, or haptic feedback on the physical environment in three dimensions that can be interacted with~\cite{schmalstieg2016,azuma1997}. This is achieved by using devices such as special headsets with see-through displays, headsets with pass-through cameras that display the real environment together with the virtual information on screens, as well as smartphones and tablets. In addition, special interaction devices may be used for haptic feedback or device-less interaction can be employed by using gestures. On a technical level, AR bridges the gap between the virtual and the real world - spatially and cognitively~\cite{schmalstieg2016}. For realizing software applications that use augmented reality, a large number of concepts need to be understood and mastered on a technical level. These include for example \emph{detectables} for guiding computer vision systems to identify real-world objects, \emph{augmentations} that stand for virtual content that is fueled into the AR environment, \emph{anchors} for specifying the positions of augmentations by connecting them to detectables, or \emph{coordinate} and \emph{transform} concepts for positioning objects in three-dimensional space~\cite{MuffF23}.
Further, context information may need to be added to the AR application, e.g. to determine the current location of a user and determine  which graphical representations should be presented~\cite{ShishkovFISV23,MuffF22}.

\section{Spatial Conceptual Modeling}
\label{sec:spatial_conceptual_modeling}

With the clarification of the terms outlined in the previous section, we can now advance to the description of the concept of \emph{spatial conceptual modeling}. Several approaches have been described in the past for combining conceptual modeling and virtual and/or augmented reality. A recent literature survey of around 200 publications revealed that especially the fields of business process management and data modeling have investigated such combinations from various perspectives, including for example the elicitation and visualization of process, data, or enterprise architecture models using augmented and virtual reality, or the three-dimensional representation of models~\cite{MuffF23bmsd}. Further approaches that were found in the survey included approaches for the model-driven engineering of virtual reality and augmented reality applications, e.g. most recently~\cite{MuffF23,MuffF21}. 

What seems to be missing so far, however, is a fundamental approach for combining the knowledge contained in conceptual models with entities in the real world on the level of metamodeling. Such an approach would permit to leverage arbitrary types of conceptual models in different languages to the spatial dimension. Although this could be done purely in formal mathematical notation, the practical benefits seem to lie in a combination with augmented reality technologies. These technologies permit to literally 'grasp' the knowledge of conceptual models in the real world. 

\begin{figure}
    \centering
    \includegraphics[width=0.65\textwidth]{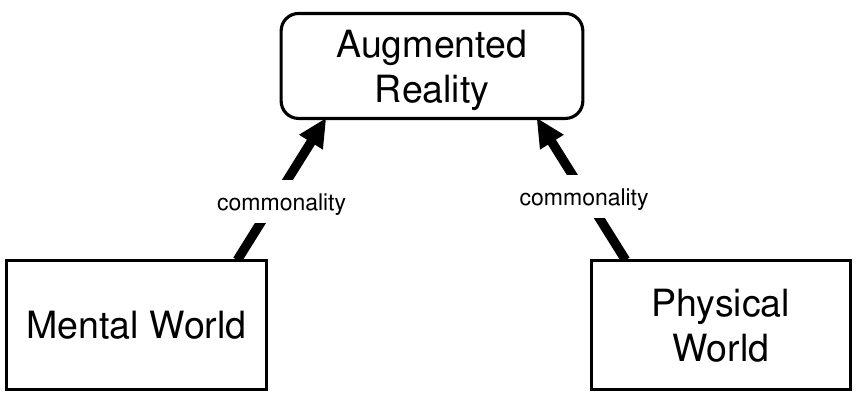}
    \caption{Augmented Reality as the Tertium Comparationis between the Mental and the Physical World}
    \label{fig:ar_tertium}
\end{figure}

We thus view augmented reality as a so-called \emph{tertium comparationis} between the mental and the physical world. The mental world relates thereby to the knowledge as expressed through conceptual models and the physical world to all entities and things sensed in the real world. Following Čyras and Lachmayer, the tertium comparationis refers to the "quality that two things that are being compared have in common"~\cite{Cyras2023}[p.15]. As depicted in Figure~\ref{fig:ar_tertium}, the commonalities of the mental and the physical world are thereby mediated via augmented reality, for example by superimposing knowledge on how to operate a machine that is part of the mental world and made explicit via a conceptual model on the actual buttons and switches of the physical machine.

Augmented reality thereby enables the anchoring of information stemming from conceptual models to entities in the real world by providing the necessary technical concepts. In terms of spatial information content it may however be needed to provide additional information, which is typically not found so far in conceptual models. This includes for example information on the location of information objects in physical space, as well as different levels of granularity in terms of different levels of detail required for the expression of knowledge in conceptual models and on the physical level - see for an example the work by Crevoiserat et al.~\cite{Crevoiserat23}. 

In a fictitious example, conceptual models could be anchored in the real world as follows. The models could contain information on the enterprise architecture of a company which is displayed in the server room when people with AR headsets enter the room. Thus, it would not be necessary anymore to operate a laptop or tablet and search for the right model. Rather, the model could be displayed once the people look at a particular server rack, e.g. to determine which applications run on that server and who is responsible for it.

We can further identify different levels of anchoring conceptual models in the real world. As shown in Table~\ref{tab:levels} five levels are proposed. First, level 0 stands for the classical, 2D-based modeling without any spatialization. Subsequently, Level 1 '\emph{Unanchored Spatial CM}' characterizes approaches where models are just presented in space, e.g. by displaying them on an AR headset or narrating their content on audio devices in space. However, on this level models are not tied to a particular physical location. This may be useful for automatically presenting models to users upon sensing and reasoning about their behavior, independent of their location. Level 2 '\emph{Model-Anchored Spatial CM}' stands for the anchoring of models to points of interest in space. For example, when a user approaches a machine at a defined physical location, a particular model with operating instructions is shown. 

\begin{table}[!t]
    \centering
    \begin{tabular}{L{1cm}L{4.6cm}p{6cm}}
        \hline\noalign{\smallskip}
         Level & Designation & Description of Anchoring Level \\
         \noalign{\smallskip}\svhline\noalign{\smallskip}
         \textbf{Level 0} & Traditional CM & No spatialization, classical 2D modeling \\
         \textbf{Level 1} & Unanchored Spatial CM & Spatialization of models without anchoring \\
         \textbf{Level 2} & Model-Anchored Spatial CM & Anchoring of models to points of interest in space\\
         \textbf{Level 3} & Statically Anchored Spatial CM & Static anchoring of model elements in space \\
         \textbf{Level 4} &  Dynamically Anchored Spatial CM & Dynamic anchoring of model elements in space\\
        \noalign{\smallskip}\hline\noalign{\smallskip}
    \end{tabular}
    \caption{Anchoring Levels of Spatial Conceptual Modeling (CM)}
    \label{tab:levels}
\end{table}

Level 3 '\emph{Statically Anchored Spatial CM}' goes one step further and refers to the anchoring of individual model elements in space. Instead of placing the whole model at some point in space, now single elements of a model can be tied to physical locations. For example, steps in the operation process of a machine may be anchored to the various switches of that machine to indicate which switch needs to be turned next. Level 4 '\emph{Dynamically Anchored Spatial CM}' extends this anchoring of model elements then in a dynamic manner. Instead of statically anchoring model elements, this may be done dynamically, e.g. based on some reasoning about the current context~\cite{MuffF22}. For example, a workflow for operating switches on different machines may be dynamically adapted based on some user action.

\subsection{Formal Characterization of Spatial Conceptual Modeling}

For realizing spatial conceptual modeling we can derive in the next step how spatial information concepts can be represented in conceptual modeling on a metamodeling level. For this purpose we will revert to the constructs of the FDMM formalism~\cite{FillRK12,FillRK12a}. FDMM has been designed to formally describe metamodels and their model instances on a technology-independent level. In contrast to other formalisms it strives for ease-of-use and simplicity in the mathematical formulation, so that also people with only little background in formal specifications can understand and apply it. 

FDMM defines metamodels $\mathbf{MM}$ as a tuple of the form $\mathbf{MM} =  \bigl\langle \mathbf{MT}, \preceq, \mathrm{domain}, \mathrm{range}, \mathrm{card}\bigr\rangle$, where $\mathbf{MT}$ stands for a set of model types. Each model type $\mathbf{MT}_i$ has in turn a tuple of object types $\mathbf{O}^T_i$, data types $\mathbf{D}^T_i$, and attributes $\mathbf{A}_i$, i.e. $\mathbf{MT}_i = \left\langle \mathbf{O}^T_i , \mathbf{D}^T_i ,\mathbf{A}_i \right\rangle$. Object types are used to both represent the types of nodes and edges in typical model diagrams or as template for arbitrary objects. Attributes can be attached to object types via the domain function. The range function determines the type of content of an attribute type - including data types or other object or model types - and the card function specifies the cardinality of attribute values in model instances. $\preceq$ is an ordering of object types for specifying inheritance relationships between object types. For the scope of this paper we omit the detailed formal relationships between the constructs of metamodels as well as the instantiation part of FDMM and refer interested readers to~\cite{FillRK12,FillRK12a} as well as further applications of FDMM in~\cite{FillHKOS13,JohannsenF15,JohannsenF17}.

In the following we will show how the information concepts from spatial computing can be mapped to metamodeling based on the outline in Section~\ref{subsec:spatial_computing} following Kuhn and Ballatore~\cite{KuhnB15}. These mappings are shown in Table~\ref{tab:adaptations}. We first consider the respective concept from spatial computing and then the corresponding FDMM concepts for spatial conceptual modeling. Finally a brief description is added.

\begin{table}[!t]
    \centering
    \begin{tabular}{L{4cm}p{3.2cm}p{4.2cm}}
        \hline\noalign{\smallskip}
         \textbf{Spatial Information Concept} &  \textbf{FDMM Concept} & \textbf{Description}\\
         \noalign{\smallskip}\svhline\noalign{\smallskip}
         Location & $\mathbf{A}_\mathrm{{coord}}$  & Set of coordinate attributes\\
                 & $\mathrm{domain}(\mathbf{A}_\mathrm{{coord}}) = \mathbf{O}^T$ & Every object type has coordinate attributes\\
                 & $\mathbf{A}_\mathrm{{transform}}$ & Set of transform attributes \\
                 & $\mathrm{domain}(\mathbf{A}_\mathrm{{transform}}) = \mathbf{O}^T$ & Every object type has transform attributes\\
        \noalign{\smallskip}\hline\noalign{\smallskip}
         Field & $\mathbf{O}^T_\mathrm{{field}}, \mathbf{A}_\mathrm{{field}}$ & Sets of field object types and attributes\\
         & $\mathrm{domain}(\mathbf{A}_\mathrm{{field}}) = \mathbf{O}^T_\mathrm{field}$ & Field attributes are assigned to field object types\\
         \noalign{\smallskip}\hline\noalign{\smallskip}
         Object & $\mathbf{a}_\mathrm{{uuid}}$ & UUID attribute for object identity\\
                & $\mathrm{domain}(\mathbf{a}_\mathrm{{uuid}}) = \mathbf{O}^T$ & Every object type has a UUID attribute\\
                & $\mathbf{O}_\mathrm{real}^{T} \subseteq \mathbf{O}^T$ & Real object types refer to objects in the real world \\
                & $\mathbf{O}_\mathrm{virt}^{T} \subseteq \mathbf{O}^T$ & Virtual object types stand for virtual objects \\
                & $(\mathbf{O}_\mathrm{real}^{T} \cup \mathbf{O}_\mathrm{virt}^{T}) \subset \mathbf{O}^T$ & The set of object types comprises all real and virtual object types  \\
                & $\mathbf{O}_\mathrm{rv}^{T} \subseteq \mathbf{O}^T$ & Object types for relating real and virtual object types \\
                & $\mathbf{A}_\mathrm{{vizrep}}$ & Set of visualization representation attributes\\
                & $\mathrm{domain}(\mathbf{A}_\mathrm{{vizrep}}) = \mathbf{O}^T$ & Every object type has a set of visualization representation attributes\\
         \noalign{\smallskip}\hline\noalign{\smallskip}
         Network & $\mathbf{O}^T$ & Nodes and edges represented as FDMM object types\\
         \noalign{\smallskip}\hline\noalign{\smallskip}
         Event & $\mathbf{O}^T_\mathrm{event}$ & Event object types for temporal events\\
         & $\mathbf{A}_\mathrm{event}$ & Attributes for events for expressing temporal properties\\
                & $\mathbf{O}^T_\mathrm{Temp}$ & Temporal event object types for temporal relations between events\\
                & $\mathbf{O}^T_\mathrm{part}$ & Participation object types for relations between events and other object types\\
        \noalign{\smallskip}\hline\noalign{\smallskip}
    \end{tabular}
    \caption{Mappings of Spatial Information Concepts to FDMM Concepts for Spatial Conceptual Modeling}
    \label{tab:adaptations}
\end{table}

Starting with the concept of \emph{Location}, this requires on the side of conceptual modeling that all modeling objects can be anchored in three-dimensional space using \emph{coordinates}. However, due to the multitude of coordinate systems necessary for spatial computing applications - e.g.\ GPS coordinates, coordinates in Building Information Modeling (BIM), coordinates of graphical devices and sensors, etc. - we subsume these under the attribute set $\mathbf{A}_\mathrm{coord}$. This set is attached to any object type using a domain function. Similarly, objects may have an orientation in space that needs to be considered and that we regard also under location. Therefore, we add a set $\mathbf{A}_\mathrm{transform}$ that holds this information. Note that also for coordinate transforms many variants exist such as rotation matrices, Euler angles, quaternions and so on. 

The concept of a \emph{Field} can also be represented in conceptual modeling via object types and assigned attributes. We denote these through $\mathbf{O}^T_\mathrm{{field}}, \mathbf{A}_\mathrm{{field}}$. In this way, for example, a temperature field could be created through the instantiation of a field object type and values from attributes assigned to it that hold e.g.\ information about the mathematical function specifying the physics of the field. Further, such a field can be combined and extended with references to other objects types as done in conceptual modeling for adding semantic information, e.g. to represent user actions that are affected by the temperature field in a laboratory environment.

For the \emph{Object} concept from spatial computing, the identity is a key property. Therefore, we foresee a distinct attribute $\mathbf{a}_\mathrm{{uuid}}$ that is used to attach a universally unique identifier to each object type. This permits the decentralized creation of new, uniquely identifiable objects across an arbitrary number of applications. We further divide the overall set of object types into object types referring to the real world $\mathbf{O}^T_\mathrm{real}$ and those standing for virtual objects $\mathbf{O}^T_\mathrm{virt}$. This can be used for example in augmented reality applications to hold information about objects in the real world, e.g.\ in the form of markers as surrogates for these objects or for identifying the position and pose of real objects using object detection techniques~\cite{Zou2023}. Virtual object types are used to represent the traditional conceptual modeling elements and relations, e.g.\ a place in a Petri net diagram, a class in UML, an inheritance relationship type in UML -- or, more advanced types such as augmentations for augmented reality applications. In addition, object types for relating real and virtual object types are foreseen, which are denoted as $\mathbf{O}^T_\mathrm{rv}$. These permit to establish connections between virtual and real world objects, e.g.\ for anchoring an augmentation with a real world object. The attributes, domain, and range functions for referencing the related real or virtual object types are omitted here for brevity. An essential aspect of conceptual models is their graphical reprsentation, even more so in three-dimensional space. Therefore, the set $\mathbf{A}_\mathrm{vizrep}$ is foreseen to include all necessary attributes related to the visual representation of object types. This may include static representations in two, three or more dimensions - e.g.\ including time and animation aspects - as well as dynamic aspects, e.g. for the realization of dynamic state changes in visualizations based on attribute states~\cite{fill2009}. 

The \emph{Network} concept of spatial computing largely corresponds to the typical constructs in conceptual models for representing mathematical graphs in various forms. Therefore, we can directly map it to FDMM object types, which are used to represent nodes and edges in graphs.

The \emph{Event} concept is used to introduce temporal aspects in spatial computing. We foresee a set of object types $\mathbf{O}^T_\mathrm{event}$ for the events themselves, a set $\mathbf{A}_\mathrm{event}$ for representing temporal properties of events in the form of attributes such as durations and time units, the object type $\mathbf{O}^T_\mathrm{Temp}$ for temporal relations between events such as before, after, etc., and the object types $\mathbf{O}^T_\mathrm{part}$ for participating relationships between events and any other object type.

Concerning the operations for spatial computing, we omit those at this stage due to the fact that FDMM does not foresee fundamental operations for models at the moment. A possible formalization could be based for example on topological relations for representing knowledge~\cite{hernandez1994}, as well as approaches for qualtitative spatial reasoning~\cite{cohn2001}. This would need to be mapped to fundamental operations on model instances such as iterations over model elements, creation, modification and deletion of elements, or operations affecting the user interface of model editors such as markings or animations.

\subsection{Possible Applications Areas of Spatial Conceptual Modeling}
\label{subsec:application_areas}

From a general perspective, spatial conceptual modeling seems most adequate for applications that combine knowledge-intensive areas with physical interactions. In such cases, knowledge needs to be elicited in the form of conceptual models, which can then either be mapped directly or via intermediate representations to the physical environment. In the following we list some potential application areas.

\textbf{Healthcare and Medical Domain}: In this domain, conceptual modeling has been used for example to represent clinical pathways, treatment processes, or regulatory processes, e.g.~\cite{BraunSBE15,FillELRLK11}. The integration of spatial aspects could help to support medical education about treatments or personnel in care delivery, where augmented reality has already been successfully investigated~\cite{munzer2019}. 

\textbf{Engineering and Manufacturing}: Several approaches have been developed for using augmented reality in this field, e.g.\ in maintenance, collaborative design, layouting, or training~\cite{regenbrecht2005}. The addition of conceptual modeling could help to deal with the challenge of integrating the large amounts of data and knowledge necessary for such applications and for making the working of AR applications more transparent. Further, spatial conceptual modeling could complement digital twin approaches in engineering by adding knowledge and process aspects to the traditionally used CAD models~\cite{albuquerque2023}.

\textbf{Physical IT Architectures and Smart Homes}: In IT and enterprise architecture scenarios, approaches have been developed to visualize systems using augmented reality~\cite{rehring2019}. Whereas also such approaches may benefit from spatial conceptual modeling through directly retrieving the necessary data from the models, we see further potential in the combination with physical IT architectures, e.g. to support maintenance, for the physical wiring and optimization of communication networks, or for end users in smart home scenarios~\cite{mahroo2019}, where the combination with conceptual models could already be successfully demonstrated~\cite{Crevoiserat23}.

\textbf{Supply Chain Management and Logistics}: For this domain, the application of augmented reality has been investigated for the optimization of business processes, e.g. for supporting the selection of next steps, for picking items, or for monitoring processes~\cite{rejeb2021,plakas2020}. However, as derived by Rejeb et al.~in a recent literature survey, several technical, organizational, and ergonomical challenges persist in this area~\cite{rejeb2021}. For this purpose, spatial conceptual modeling could help to reduce the complexity of integrating augmented reality technologies in such processes and increase the transparency of according applications. This could support both the acceptance of the technologies by the involved workers - e.g. for addressing privacy issues and fear of control - as well as to ease the alignment of the technology with organizational requirements.

\textbf{Civil Engineering, Smart Cities, and Cultural Heritage}: At the core of these areas are today large amounts of information that need to be processed electronically. May it be Building Information Models (BIM) that are used for the engineering and operation of buildings, data for environmental optimizations and sustainability, legal and compliance aspects, or historical knowledge about cultural artifacts. The use of spatial conceptual modeling could aid here in aligning the vast amounts of knowledge with the diverse sources of spatial data and presenting them through augmented reality applications~\cite{maietti2021,kim2017,MuffFKK22}. Examples would include the integration of knowledge about cultural heritage and spatial architectural data, as well as data for environmental and sustainability purposes, e.g. in smart city planning or for informing citizens about environmental-friendly behavior in the built environment.

In summary, the provision of a generic approach for spatial conceptual modeling that is applicable to diverse application areas would not only permit the easier exchange of knowledge on implementing augmented reality applications. It would also support these areas with a common technical approach for integrating data and knowledge, thus potentially making the implementation more efficient and effective.

\section{Conclusion and Outlook}
\label{sec:conclusion}

In this paper, a first outline of the approach of spatial conceptual modeling has been presented, including an intial formal characterization. At its core, spatial conceptual modeling aims to bridge the gap between the mental and physical world through augmented reality applications in terms of knowledge aspects. There are however a number of open issues that will need to be addressed. This concerns both conceptual and implementation aspects of spatial conceptual modeling. For example, it will need to be investigated, whether existing metamodeling platforms can be used for its realization in terms of required concepts and technologies~\cite{KaragiannisK02}. Further, it will need to be evaluated whether the wide range of existing conceptual modeling approaches is adequate for transitioning them to the spatial environment or which changes are required~\cite{karagiannis2022domain}. Finally, upcoming technologies in artificial intelligence could be used for easing the creation of spatial representations of model elements, which requires otherwise specialized knowledge about three dimensional modeling~\cite{FillMuff2023,FillFK23}. Last but not least, business models and further use cases for spatial conceptual modeling need to be designed and evaluated, e.g. in the context of metaverse-like environments.

\bibliographystyle{spmpsci}
\bibliography{literature}

\end{document}